\documentclass[sigconf]{acmart}
\usepackage{lipsum}
\usepackage{booktabs}
\usepackage{pifont}
\usepackage{balance}

\AtBeginDocument{%
  }

\copyrightyear{2026}
\acmYear{2026}
\setcopyright{rightsretained}
\acmConference[RecSys '26]{Proceedings of the Twentieth ACM Conference on Recommender Systems}{September 28--October 2, 2026}{Minneapolis, Minnesota, USA}
\acmBooktitle{Proceedings of the Twentieth ACM Conference on Recommender Systems (RecSys '26), September 28--October 2, 2026, Minneapolis, Minnesota, USA}

\begin{document}
\emergencystretch 3em
\title[Beyond Co-purchase Relation: Evolution of Complementary Recommendations at Allegro]{Beyond Co-purchase Relation: \\ Evolution of Complementary Recommendations at Allegro}







\author{Aleksandra Osowska-Kurczab}
\authornote{All authors contributed equally to this research.}
\orcid{0000-0001-5764-522X}
\email{aleksandra.kurczab@allegro.com}
\affiliation{%
  \institution{Allegro.com}
  \country{Poland}
}

\author{Klaudia Nazarko}
\authornotemark[1]
\orcid{0009-0009-4133-211X}
\email{klaudia.nazarko@allegro.com}
\affiliation{%
  \institution{Allegro.com}
  \country{Poland}
}

\author{Eliška Kosturová}
\authornotemark[1]
\orcid{0009-0004-2333-7367}
\email{eliska.kosturova@allegro.com}
\affiliation{%
  \institution{Allegro.com}
  \country{Czech Republic}
}

\author{Lidia Wojciechowska}
\orcid{0009-0009-5207-681X}
\affiliation{%
  \institution{Allegro.com}
  \country{Poland}
}

\author{Michał Bień}
\authornote{Work done at Allegro.com.}
\orcid{0000-0002-1269-0163}
\affiliation{%
  \institution{NVIDIA}
  \country{Poland}
}

\renewcommand{\shortauthors}{Osowska-Kurczab et al.}

%
\begin{abstract}
When a customer adds a professional camera to their cart, should the system suggest a matching lens, a generic tripod, or another camera body? Complementary Product Recommendation is vital for comprehensive basket building, yet standard models often fail to distinguish between items that are merely bought together and those that truly work together. In this paper, we present AlleCompanion: a production-scale retrieval framework deployed at Allegro.com that transforms noisy behavioural signals into precise semantic compatibility. We mitigate the intrinsic noise in large-scale co-purchase traffic by combining data-level filtering heuristics with a category-constrained Two Tower architecture. Within this framework, the Category Adapter guides the model in the embedding space, constraining candidates within logically complementary boundaries. Since modelling authentic user behaviour at scale is inherently difficult, we introduce ComCat, a multi-source Complementary Categories Mapping. ComCat acts as a translational layer that distils meaningful patterns from noisy traffic into a maintainable and controllable solution, integrating expert rules, human-in-the-loop feedback, LLM-based reasoning, and statistical mining. Our experimental results demonstrate that combining explicit category-level constraints with neural architectures effectively filters out co-purchase noise to surface recommendations that satisfy real-world user needs. Serving over 20 million active users monthly, the framework delivers significant uplifts in attributed GMV for organic discovery and drives substantial revenue growth in sponsored placements.
\end{abstract}


  


\maketitle

\section{Introduction}
\label{sec:introduction}
The objective of modern e-commerce recommendation systems has evolved from isolated item discovery towards modelling user’s comprehensive purchase intent. While similarity-based models excel at selecting specific items (such as suggesting another smartphone), business value is often driven by complementarity. For example, recommending a protective case or a high-speed charger to a phone is what ultimately maximises shopping basket value and customer satisfaction. However, modelling complementarity presents a distinct challenge compared to similarity; it requires a fundamental shift from matching items with overlapping features (\textit{substitutes}) to identifying items that work well together (\textit{complements}).

Every month, over 20 million active buyers visit Allegro to explore a vast catalogue of products from more than 150 thousand sellers. Within this massive ecosystem, complementary recommendations stand out as one of the most profitable discovery mechanisms \cite{osowska_suggest_2025}, driving engagement for both organic and sponsored content. We formulate these recommendations as a product-to-product relation where a target item functionally extends a query product. In practice, users frequently acquire these items together, a co-purchasing behaviour that is strongly encouraged by platform incentives to source multiple goods from the same seller. However, discovering genuine complementary pairs within historical transactions requires navigating complex, cross-category relationships bound by strict item-level compatibility --- a fundamental shift from merely matching items with overlapping features.

Translating this goal into an accurate complementary retrieval system poses three distinct challenges. Firstly, recommendations must bridge the gap between broad category-level associations and precise item-level compatibility. This requires models to understand which product groups are complementary while strictly enforcing the compatibility on item-specific attributes like brand, model or size. Secondly, the modelled relations must generalize to cold-start and long-tail items while respecting the inherent asymmetry of complementary pairs --- for instance, while a phone charger complements a smartphone, a smartphone does not complement a charger. Finally, learning from historical co-purchase data is severely hindered by noise \cite{mane_quadruplet_2019}. While transaction logs capture historical co-purchase behaviour, they frequently combine distinct intents. For example, a user purchasing multiple flavours of dog food, or both dog and cat food, represents a desire for substitute items or entirely separate needs rather than true functional complements. Overcoming this requires logical guidance beyond mere transaction frequency.

To address these challenges, we introduce AlleCompanion, an end-to-end framework for complementary product recommendation. Our work encompasses the complete lifecycle of the recommendation system, from architectural design, dataset construction and offline ablations to successful production deployment. The main contributions of this work are summarized as follows:
\begin{itemize}
    \item \textbf{AlleCompanion}, Two Tower architecture featuring category-conditioned retrieval. It steers the latent space towards complementary categories while maintaining precise item-level alignment, bridging the gap between broad category-level associations and compatibility.
    \item \textbf{ComCat}, a multi-source mapping integrating expert rules, LLM insights, and statistical mining to guide the model's responses towards logically complementary relations.
    \item An empirical study on the \textbf{impact of various dataset definitions}, providing critical insights into the practical challenges of filtering out non-complementary purchase intents and isolating true item-level compatibility.
    \item \textbf{Lessons from offline ablations and online A/B testing}, demonstrating the model’s practical value and performance gains for both organic and sponsored recommendations.
\end{itemize}

\section{Related Work}
\label{sec:related-work}

E-commerce marketplaces drive basket growth through financial incentives (e.g. free delivery thresholds \cite{sun_revisiting_2022}), strategic UX placement and gamification mechanisms \cite{bayir_gamification_2024}. These elements are injected across the user journey --- from ``complete the look'' suggestions to value-added services at checkout. To scale these strategies, platforms rely on complementary recommendation algorithms \cite{li_complementary_2024} that identify relevant items across diverse contexts.

However, mining these relationships is complex, as raw co-purchase logs often fail to distinguish between joint demand and mere alternatives \cite{sugahara_is_2024, hao_pcompanion_2020}. Recent research introduces distant supervision methods to filter this noise, such as excluding co-purchase pairs with high co-view overlap \cite{hao_pcompanion_2020} or applying a substitute penalty \cite{zhao_recommending_2017}. While these heuristics improve signal quality, they often struggle with the nuances of complementarity or become too complex to maintain in production. We address this trade-off by refining behavioural filters and evaluating diverse data sources beyond simple co-purchase traffic.

In addition to behavioural logs, complementarity requires understanding compatibility between products through Knowledge Graphs \cite{zalmout_all_2021} or Large Language Models (LLMs) \cite{jeon_leveraging_2025}. LLMs, acting as repositories of world knowledge, can generate explanations \cite{li_explainable_2024}, identify complementary concepts \cite{huang_ccgen_2023} or act as few-shot annotators \cite{yamasaki_knowledgeaugmented_2025, jeon_leveraging_2025}. For platforms with sparse interaction logs, transferring such knowledge into a universal embedding space is crucial for identifying relations across the long tail items \cite{papso_complementary_2023}. We extend these methodologies through ComCat, a multi-source category mapping that distils LLM-based reasoning and expert logic into a controllable translational layer for consistent cross-category retrieval.

The architectural landscape for complementary tasks has evolved from content-based similarity \cite{mane_quadruplet_2019, xu_knowledgeaware_2020} towards sequential modelling. GNNs have become the standard for capturing non-transitive dependencies \cite{chen_enhanced_2023, luo_spectral_2024}, while transformer-based architectures \cite{mokhtari_trex_2026} address the temporal dynamics of basket building \cite{zhang_learning_2022}. Current research focuses on hybrid frameworks that fuse GNN-derived structural knowledge with multi-modal content \cite{wang_multimodal_2025}. A prominent example is P-Companion \cite{hao_pcompanion_2020}, which utilises specialised embedding spaces to balance relevance with diversity \cite{yan_personalized_2022}. Yet, while such hybrid systems achieve high precision, they often introduce significant computational overhead --- particularly when involving real-time GNN inference --- complicating real-time retrieval at scale. AlleCompanion builds upon the hybrid paradigm established by P-Companion, but prioritises architecture and maintenance simplicity. 

\section{Methods}
\label{sec:methods}

\subsection{Architecture Definition}
\label{sec:methods:architecture-definition}
\begin{figure*}
    \centering
    \includegraphics[alt={A system diagram of the AlleCompanion Two Tower neural network architecture. It shows a shared product encoder mapping query (x_q) and target product features into embeddings. The query tower features a Category Adapter that concatenates the base query embedding (v_q) with a complementary category embedding (e_comp) to produce the final query embedding (q). The network optimizes a joint loss function combining a retrieval loss (L_retrieval) and an auxiliary category reconstruction loss (L_aux)."},width=0.9\linewidth]{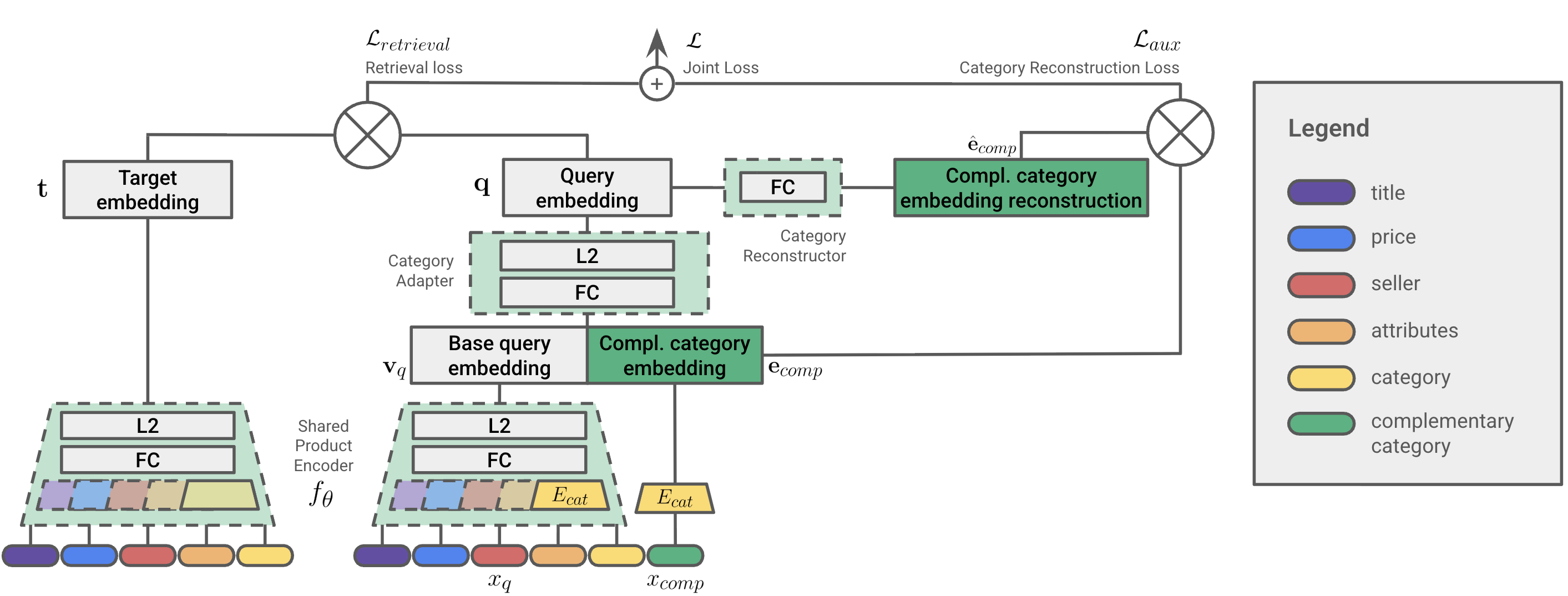}
    \caption{AlleCompanion architecture. Two Tower model extended with Category Adapter and Category Reconstruction Loss.}
    \label{fig:allecompanion-architecture}
\end{figure*}

AlleCompanion is a content-based model designed specifically for complementary product recommendation. Our proposed architecture features a Category Adapter, which projects the item embedding into a complementary category latent space. Furthermore, we utilise a category reconstruction loss \cite{category_reconstruction} that enhances the model's ability to learn robust representations of these categories. The high-level architecture of AlleCompanion is illustrated in Figure \ref{fig:allecompanion-architecture}.

\subsubsection{Two Tower}
The proposed approach utilises a Two Tower deep learning model, which maps query and target product features into a shared embedding space. The training objective is to maximise the inner product between the query and target vector representations. We frame this retrieval task as a classification problem, leveraging a sampled softmax loss function ($\mathcal{L}_{retrieval}$) \cite{yt-dnn-recommender} integrated with a mixed negative sampling strategy \cite{mixed-negative-sampling}, and temperature scaling \cite{temperature_scaling}. To improve convergence and ensure numerical stability, we implement parameter sharing between the query and target towers. This unified module is designated as the Product Encoder.
Under the hood, the architecture consists of dedicated embedding tables that transform each individual item feature into a low-dimensional vector. Once concatenated, these vectors are processed through a multi-layer perceptron (FC) and L2-normalised to produce the final representation.

The model is trained in an item-to-item regime, learning the relationship between products co-purchased by a user within a short time window. To account for the volatility of a million-scale product catalogue, we represent each item via its content features, rather than ID features. In Vanilla Two Tower, by default, a product is represented using its title, price, and category features. This content-based approach allows for the seamless integration of additional textual and categorical features, such as product attributes and seller IDs, respectively. To capture broader structural relationships, the category feature is derived from the hierarchical taxonomy of categories maintained in Allegro's product catalogue (e.g., \textit{Allegro > Electronics > Smartphones)}.

\subsubsection{Category Adapter}
While a standard Two Tower architecture primarily focuses on identifying similar items in a shared embedding space, our goal is to explicitly direct the model towards retrieving complementary products. To achieve this, we introduce a Category Adapter module that constrains the retrieval process to a requested complementary category.
Firstly, the query product features $x_q$ are passed through the Shared Product Encoder $f_{\theta}$ to obtain the base query embedding $\mathbf{v}_{q} = f_{\theta}(x_q)$. In parallel, the target complementary category $x_{comp}$ is projected into a dense vector $\mathbf{e}_{comp} = \mathbf{E}_{cat}(x_{comp})$, utilising the shared category embedding table $\mathbf{E}_{cat}$ from the Product Encoder. Lastly, these two vectors are concatenated and passed through a FC layer (weights $\mathbf{W}$ and bias $\mathbf{b}$) with normalisation to produce the final query representation:
$$\mathbf{q} = \text{Norm}(\mathbf{W} [\mathbf{v}_{q} \parallel \mathbf{e}_{comp}] + \mathbf{b})$$
Category Adapter is integrated into the query tower, where the product representation $\mathbf{v}_{q}$ is augmented with the target complementary category $\mathbf{e}_{comp}$. During training, this category is dynamically derived from the ground-truth target item to ensure that the model learns to associate specific products with their appropriate complementary classes.

\subsubsection{Category Reconstruction Loss}
To guide the Category Adapter towards more discriminative signals, we introduce an auxiliary reconstruction loss $\mathcal{L}_{aux}$. Specifically, we project the final query embedding $\mathbf{q}$ through a single FC layer to reconstruct the complementary category representation:
$$\hat{\mathbf{e}}_{comp} = \mathbf{W}_{aux}\mathbf{q} + \mathbf{b}_{aux}$$
We optimise this alignment using a sampled softmax loss, effectively minimising the distance between the projected query $\hat{\mathbf{e}}_{comp}$ and the true complementary category embedding $\mathbf{e}_{comp}$.

The final optimization objective $\mathcal{L}$ is formulated as a joint loss, combining the primary retrieval loss $\mathcal{L}_{retrieval}$ and the auxiliary reconstruction loss:$$\mathcal{L} = \mathcal{L}_{retrieval} + \mathcal{L}_{aux}$$

\subsection{Dataset Construction}
\label{sec:methods:dataset-construction}
Co-purchase signals serve as the primary basis for capturing latent complementary relations between products during the training of AlleCompanion. While they serve as a strong proxy for complementarity, raw transactional data is inherently noisy and often fails to distinguish true complementary items from substitutable ones. To address that, we refine the data through filtering and heuristics.

\subsubsection{Raw dataset generation}
We define a co-purchase session as a sequence of items $(i_1, i_2, \dots, i_n)$ purchased by a single user within a specific time window. From these sessions, we derive ordered asymmetric pairs $(i_j, i_k)$ where $j < k$, ensuring the sequence models the directionality of complementary needs.

The quality of these generated pairs depends heavily on the temporal window used to define co-occurrence. Empirical observations of Allegro traffic suggest that while same-cart purchases provide high frequency, they are often dominated by highly similar or identical items. Shifting focus towards longer session windows significantly increases item diversity and volume while preserving semantic relevance; by contrast, much longer windows introduce excessive noise from unrelated purchases. To address this trade-off, the exact session window duration was selected via hyperparameter tuning evaluated against a holdout validation set, balancing semantic connection and item diversity.

\subsubsection{Behavioural filtering and heuristics}
To ensure that the model learns from representative user behaviour rather than unrelated interactions or outliers, a multi-step filtering process is applied. First, ``heavy buyers'' whose transaction volume exceeds the 99th percentile are excluded to prevent high-frequency outliers from biasing the learned distribution. This is followed by heuristic rules, designed to distinguish true complementary pairs from substitutes or unrelated associations, focusing on two primary signals: pair count (the frequency of an item pair across sessions) and category alignment (categorical overlap between the items’ respective departments and categories). 

Following established practices \cite{hao_pcompanion_2020}, we conducted an internal annotation task involving 400 item pairs sampled from the co-purchase dataset. Annotators assigned relationship labels—\textit{complementary}, \textit{substitutable}, or \textit{unrelated}—requiring a consensus of two per pair. Analysis showed that enforcing a minimum pair count effectively filters niche, coincidental co-purchases, reducing the presence of unrelated items from 44\% to 22\%. However, excessively high thresholds predominantly isolate substitutes (increasing from 28\% to 43\%) rather than complements (which only rise from 29\% to 36\%).

To rebalance the dataset toward a target hierarchy of complementary $>$ substitutes $>$ unrelated, a category alignment heuristic was implemented. This heuristic requires product pairs to share the same department but belong to different categories (e.g., \textit{Tripod} and \textit{Camera Lens} within \textit{Electronics}). Applying this constraint filtered the baseline pool ---originally consisting of 36\% complementary, 16\% substitutes, and 48\% unrelated pairs --- into a rebalanced dataset comprising 61\% complementary, 4\% substitutes, and 35\% unrelated pairs. The resulting structural bias toward complementary relationships aligns with established patterns showing that such pairings frequently span distinct product types \cite{hao_pcompanion_2020}.

All filters were applied to the testset to ensure consistency, with the exception of the minimum pair count. The exclusion of this specific threshold allows for the evaluation of model performance across a broader distribution of real-world traffic. 

\subsection{Online Deployment}
\label{sec:methods:online-deployment}
Our online deployment architecture follows a standard embedding-based retrieval paradigm. The AlleCompanion model is periodically retrained on a single NVIDIA T4 (16GB) GPU, and the Approximate Nearest Neighbour index is refreshed daily using the Faiss library for efficient serving. This setup enables real-time recommendation retrieval with strict millisecond-level latency guarantees. In production, AlleCompanion serves as one of the retrievers for the recommendation carousel. It is deployed alongside collaborative filtering methods and is supplemented by heuristic fallbacks, such as category bestsellers, to ensure high coverage.

During online inference, explicit target items are naturally unavailable. To address this, we utilise a predefined complementary mapping to determine the valid set of target categories for a category of a given query. The model processes the query features and integrates the target category as a conditioning signal, dynamically projecting the query into a specific latent subspace \cite{hao_pcompanion_2020}. Finally, the candidate groups retrieved from each respective target category are interleaved to diversify the output and enhance the visual presentation of the recommendation carousel. Decoupling the model from ComCat enables updates to complementarity rules without model retraining.

\subsection{Complementary Categories Mapping}
\label{sec:methods:comcat}
Determining the optimal target category for a given query item is a non-trivial challenge that requires balancing relevance with diversity to satisfy varied user needs. To address this, we use ComCat as a robust proxy for ground-truth complementarity. This approach ensures the model retrieves logical results even for cold-start products or items with low traffic coverage, where historical behavioural signals are often missing or unreliable.

\subsubsection{Sources of complementary categories}
To achieve both high precision and broad catalogue coverage, we utilise a heuristic ensemble that merges three distinct data sources, weighing expert domain knowledge against raw behavioural noise. Crucially, the mapping is modelled as a directed relationship, capturing the asymmetry of complementary categories (e.g., a primary purchase driving the need for an accessory).

\textbf{Automated Co-Purchase Heuristics:}
This source identifies relationships by mining categorical co-occurrence patterns within purchase sessions. For each category, we generate directional permutations of items purchased within the same session—excluding high-volume outliers—to capture asymmetric category pairs (e.g., \textit{Smartphone} $\rightarrow$ \textit{Case}). The strength of these pairs is quantified using Jaccard similarity to ensure the signal is driven by specific co-purchase intent rather than global popularity. To prune accidental associations, we apply structural filters based on taxonomic tree distance, periodicity signals, and price ratio constraints.

\textbf{Human-in-the-Loop Annotations:}
To address coverage gaps for cold-start and low-traffic categories, we integrated a human-in-the-loop component focused on less obvious pairings. Product pairs were sourced either from filtered co-purchase traffic or through an LLM-assisted workflow \cite{huang_ccgen_2023} that identifies complementary concepts. Expert annotators labelled these pairs as \textit{complementary}, \textit{substitutable}, or \textit{unrelated}. Complementary signals were then aggregated to the category level to provide a reliable ground truth.

\textbf{Rule-Based Expert Logic:}
We employ an expert-driven source to capture compatibility aspects that behavioural data might overlook. Business experts define high-precision rules based on fine-grained product parameters to ensure strict technical alignment between recommended items. Once aggregated to the category level, these rules provide a compatibility-enabled mapping entirely independent of purchase traffic.

\subsubsection{Integration of the sources}
\label{sec:methods:comcat:merging}
The final mapping is constructed by merging these sources into a prioritised hierarchy based on expert assessment: Annotations, followed by Rule-Based logic, and finally Automated Heuristics as a broad-reach fallback. To adapt to specific business needs, this mapping can be extended with a same-category rule that maps a category to itself, such as headphones to headphones, thereby supporting the mining of alternative products. This configuration allows a single AlleCompanion instance to mix both complements and substitutes, without the overhead of maintaining multiple specialised models.

\section{Results}
\label{sec:results}
We evaluate our framework through offline experiments and online A/B tests. Our offline analysis includes an ablation study of the AlleCompanion architecture, a summary of insights gained from complementary dataset construction, and an assessment of the ComCat mechanism. Finally, we report the performance gains observed during online A/B testing, focusing on key business metrics.

\begin{figure}[H]
    \centering
    \includegraphics[alt={A visual comparison of recommendation results from three models given a green Apple iPhone as the query product. The Vanilla Two Tower model incorrectly outputs alternative smartphone bodies of different colors. The Seller Two Tower with Hard Filtering outputs phone cases, but they are physically incompatible with the query model. The AlleCompanion framework successfully retrieves a diverse selection of compatible phone cases tailored precisely to the query product's form factor.},width=0.98\linewidth]{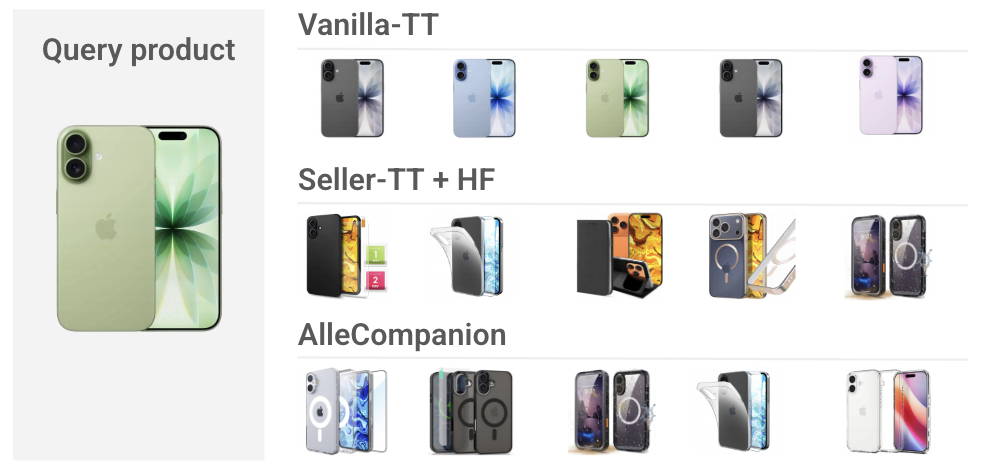}
    \caption{Recommendations generated by Vanilla-TT, Seller-TT + HF and AlleCompanion models.}
    \label{fig:candidates}
\end{figure}

\begin{table*}
    \centering
    \small 
    \Description[alttext]{alttext}
    \caption{Offline results of complementary recommendations with AlleCompanion benchmarked against other content-based models. Best results are bolded.}
    \begin{tabular}{cccccccc}
         \toprule
          & \multicolumn{5}{c}{\textbf{Evaluation: Test}} & \multicolumn{2}{c}{\textbf{Evaluation: ComCat}} \\
         \cmidrule(lr){2-6} \cmidrule(lr){7-8}
         \textbf{Model} & \textbf{Recall@20} $\uparrow$ & \textbf{MRR@20} $\uparrow$ & \textbf{Target Cat. Cons.} $\uparrow$ & \textbf{Seller Cons.} $\uparrow$ & \textbf{Attr. Cons.} $\uparrow$ & \textbf{Recall@20} $\uparrow$ & \textbf{MRR@20} $\uparrow$ \\
          \midrule
         Vanilla-TT & 0.0447 & 0.0103 & 0.0654 & 0.2237 & \textbf{0.4225} & - & - \\
         Seller-TT & 0.1219 & 0.0286 & 0.0967 & \textbf{0.9309} & 0.3309 & - & - \\
         Seller-TT + HF & 0.3782 & 0.1789 & \textbf{1.0000} & 0.8334 & 0.2320 & 0.0451 & 0.0177 \\
         AlleCompanion w/o Attr & 0.4537 & 0.1947 & 0.8283 & 0.6605 & 0.2155 & 0.0913 & 0.0244 \\
         AlleCompanion & \textbf{0.4567} & \textbf{0.2000} & 0.8267 & 0.6619 & 0.2181 & \textbf{0.0952} & \textbf{0.0258} \\
         \bottomrule
    \end{tabular}
    \label{tab:offline-results-models}
\end{table*}

\subsection{Architecture Definition}
\label{sec:results:architecture-definition}
The evaluation of AlleCompanion is strictly guided by our core design objectives. Specifically, the model must effectively learn from co-purchases to retrieve highly relevant items, explicitly generate candidates from targeted complementary categories, and promote same-seller recommendations to incentivise joint purchases. Furthermore, it must maintain a strong awareness of fine-grained product features essential for strict compatibility.

\subsubsection{Metrics}
 To assess retrieval performance, we utilise standard Recall@k and MRR@k metrics. To provide a more qualitative analysis of the model's adherence to compatibility and platform incentives, we further evaluate the following consistency metrics:
\begin{itemize}
    \item \textbf{Target Category Consistency} (Target Cat. Cons.): The average proportion of retrieved candidates that correctly belong to the intended complementary target category, evaluating the model's adherence to categorical mapping.
    \item \textbf{Seller Consistency} (Seller Cons.): The average proportion of recommended candidates offered by the exact same-seller as the query product, quantifying the model's effectiveness in encouraging single-parcel deliveries.
    \item \textbf{Attribute Consistency} (Attr. Cons.): The average overlap ratio of matching features (e.g., brand, model) between the query item and the retrieved candidates, serving as a proxy for strict item-level compatibility.
\end{itemize}

\subsubsection{Model variants \& evaluation protocol}
To evaluate the effectiveness of our proposed architecture, we benchmark AlleCompanion against several baselines rooted in the standard Two Tower framework. This allows us to isolate the impact of our specific architectural contributions. We compare the following model configurations:
\begin{itemize}
    \item \textbf{Vanilla Two Tower (Vanilla-TT)}: A standard dual-encoder trained on the transactional dataset described in Section \ref{sec:methods:dataset-construction}, serving as a baseline for capturing general co-purchase relations.
    \item \textbf{Seller Two Tower (Seller-TT)}: An extension of the vanilla baseline that incorporates the seller ID feature, explicitly designed to encourage the model to generate same-seller recommendations.
    \item \textbf{Seller-TT with Hard Filtering (Seller-TT + HF)}: The Seller-TT model augmented with a post-processing heuristic. This baseline explicitly retrieves items from the complementary category by applying a same-seller hard filter to the candidates list after generation (300 candidates).
    \item \textbf{AlleCompanion (AlleCompanion w/o Attr)}: Our proposed architecture, trained with the seller ID feature. Unlike the hard-filtering baseline, this model utilises the Category Adapter to intrinsically guide candidate generation towards the target complementary categories directly within the latent space.
    \item \textbf{AlleCompanion}: The final version of our proposed model that incorporates fine-grained product attributes, extending its ability to enforce item-level compatibility constraints.
\end{itemize}

To ensure fair and consistent comparison, identical hyperparameters --- determined through an independent tuning phase --- were applied across all configurations. The models were trained using the AdamW optimizer, with the training process incorporating temperature scaling and mixed negative sampling. Training was conducted on a co-purchase dataset spanning 90 days of transaction data, refined through the filters and heuristics described in Section \ref{sec:methods:dataset-construction} (see Table \ref{tab:dataset-statistics} for detailed statistics). Performance was then evaluated on a subsequent 7-day holdout test set to ensure temporal separation. Two primary scenarios were employed:
\begin{itemize}
    \item \textbf{Test}: A standard evaluation utilising ground-truth target categories from the holdout set
    \item \textbf{ComCat}: Evaluation using the generated category mapping (Section \ref{sec:methods:comcat}) to simulate online deployment where the ground-truth target category is unknown (Section \ref{sec:methods:online-deployment}).
\end{itemize}

\subsubsection{Discussion}
As observed in the results in Table \ref{tab:offline-results-models}, relying solely on the Vanilla-TT model to capture complementary relations leads to suboptimal retrieval performance, a degradation likely attributable to the high volume of noise inherent in raw co-purchase data. However, incorporating the seller feature (Seller-TT) proves highly beneficial, as it significantly strengthens the modelling of co-purchase dynamics. This improvement is largely driven by the high prevalence of same-merchant transactions in our data, with 59\% of co-purchase pairs in the test set belonging to the same-seller. By explicitly capturing this signal, the model not only aligns with historical user behaviour but also substantially improves seller consistency in its recommendations.

While applying a hard post-filter (Seller-TT + HF) enforces target category alignment and boosts retrieval metrics, the approach is fundamentally unscalable. Even when oversampling an initial candidate pool 15 times larger than the target list size, the median number of successfully retrieved candidates decreased to 7, highlighting the severe inefficiency of post-hoc filtering.

In contrast, our proposed AlleCompanion w/o Attr architecture demonstrates substantial improvements in both Recall@20 and MRR@20. Inherently, by guiding candidate generation, it ensures high target category and seller consistency without the scalability bottlenecks of post-filtering. Finally, augmenting this architecture with fine-grained product attributes (AlleCompanion) yields the strongest overall performance, boosting both relevance metrics and attribute consistency. This enhances the model's capacity to recommend not only complements but also compatible items.

We observe that attribute consistency is highest for the baseline Two-Tower models; this is expected, as their primary objective is to retrieve similar items with overlapping features. Conversely, AlleCompanion shifts its focus towards cross-category compatibility, yielding an attribute consistency of ~21.8\%. Importantly, this aligns closely with the ground-truth distribution of the holdout test set, where organic co-purchases exhibit an empirical baseline of 21\% attribute consistency.

Figure \ref{fig:candidates} provides a qualitative comparison of the candidates generated by the Vanilla-TT, Seller-TT + HF, and AlleCompanion models. In this example, the query item is an \textit{Apple iPhone 17}, and the target complementary category is \textit{Phone Cases and Covers}. The Vanilla-TT model focuses exclusively on the similarity aspect, primarily retrieving other iPhones that differ only in colour or built-in memory configuration. While the Seller-TT + HF model's candidates are successfully restricted to the target category, it fails to maintain strict compatibility, proposing cases for the \textit{iPhone 17 Pro} (featuring three lenses instead of two). In contrast, the AlleCompanion model provides a diverse selection of \textit{iPhone 17} cases, demonstrating its ability to simultaneously respect category constraints and fine-grained compatibility requirements.

\subsection{Dataset Construction}
\label{sec:results:dataset-construction}
\begin{table*}[t]
    \centering
    \caption{Statistics for training configurations corresponding to 90 days of purchase activity.}
    \begin{tabular}{lrrr}
    \toprule
    \textbf{Dataset} & \textbf{Trainset} & \textbf{Unique items} & \textbf{Unique categories} \\
    \midrule 
        Transactions & 3,611,848 & 779,199 & 7,555 \\
        Filtered Transactions & 829,841 & 326,238 & 3,004 \\
        Expert Rules & 3,015,027 &  1,446,727 & 5,268 \\
        Expert Rules + Same-Seller & 2,553,622 & 1,029,244 & 5,131 \\
    \bottomrule
    \end{tabular}
    \label{tab:dataset-statistics}
\end{table*}
While heuristics (Section \ref{sec:methods:dataset-construction}) increase share of complementary pairs, they don't narrow down to precise item-level compatibility across specific attributes like brand, model or size. We hypothesized that integrating expert domain knowledge can shift the focus from broad category logic to strict technical alignment, capturing the underlying reasons \textit{why} products truly complement one another.

\subsubsection{Dataset variants \& evaluation protocol}
To evaluate the impact of this expert knowledge on model performance, the Transactions dataset (Section \ref{sec:methods:architecture-definition}) is compared against three variants derived from rule-based expert logic (Section \ref{sec:methods:comcat}). These variants are designed to explore the trade-offs between data volume and precision (Table \ref{tab:dataset-statistics}):

\begin{itemize}
    \item \textbf{Transactions}: The primary dataset derived from co-purchase traffic, utilising the behavioural filters and heuristics previously described.
    \item \textbf{Filtered Transactions}: A subset of the Transactions dataset restricted to pairs that strictly match expert compatibility rules. While this produces a high-precision set of complementary pairs, it results in a 77\% reduction in total data volume.
    \item \textbf{Expert Rules}: To maintain volume while minimising noise, synthetic pairs were generated using category and attribute constraints derived from rule-based expert logic. To ensure the relevance of the resulting pairs, the dataset was restricted to active products with high user engagement and constrained by price-proximity rules.
    \item \textbf{Expert Rules + Same-Seller}: A variant of the Expert Rules dataset with an additional same-seller filter applied to align the training distribution with business requirements.
\end{itemize}

\begin{table*}[t]
    \centering
    \small 
    \Description{A comparative breakdown of performance metrics across Test and ComCat evaluation sets for different training strategies.}
    \caption{Performance of the AlleCompanion model across various training configurations. Best and second-best results are bolded and italicized, respectively.}
    \begin{tabular}{ll ccc ccc}
        \toprule
         & & \multicolumn{3}{c}{\textbf{Evaluation: Test}} & \multicolumn{3}{c}{\textbf{Evaluation: ComCat}} \\
         \cmidrule(lr){3-5} \cmidrule(lr){6-8}
         \textbf{Pretrain} & \textbf{Finetune} & \textbf{Recall@20} $\uparrow$ & \textbf{MRR@20} $\uparrow$ & \textbf{Attr. Cons. $\uparrow$} & \textbf{Recall@20} $\uparrow$ & \textbf{MRR@20} $\uparrow$ & \textbf{Attr. Cons. $\uparrow$} \\
         \midrule
         -- & Transactions& \textbf{0.4567} & \textbf{0.2000} & 0.2181 & \textbf{0.0957} & \textbf{0.0260} & 0.2411 \\
         -- & Filtered Transactions & 0.3222 & 0.1221 & \textbf{0.2433} & 0.0801 & 0.0213 & \textbf{0.2541} \\
         \midrule
         -- & Expert Rules & 0.0818 & 0.0243 & 0.1815 & 0.0127 & 0.0035 & 0.2070 \\
         -- & Expert Rules + Same-Seller& 0.2717 & 0.0960 & 0.2017 & 0.0471 & 0.0124 & 0.2288 \\
         \midrule
         Expert Rules & Transactions & 0.4458 & 0.1932 & 0.2183 & 0.0934 & 0.0253 & 0.2404 \\
         Expert Rules + Same-Seller& Transactions & 0.4429 & 0.1918 & 0.2191 & 0.0926 & 0.0252 & 0.2420 \\
         \bottomrule
    \end{tabular}
    \label{tab:offline-results-datasets}
\end{table*}

Experiments employ the AlleCompanion architecture defined in Section \ref{sec:methods:architecture-definition}, utilising the model parametrization and evaluation protocol detailed in Section \ref{sec:results:architecture-definition}.

\subsubsection{Discussion} As illustrated in Table \ref{tab:offline-results-datasets}, the base Transactions dataset remains superior in terms of Recall@20 and MRR@20 across both evaluation scenarios, which is explained by its close alignment with the underlying test distribution. However, a notable trade-off is observed with the Filtered Transactions variant. While this configuration drops in relevance metrics, it provides the highest absolute boost to attribute consistency. This suggests that although expert filters restrict the model's discovery range, they successfully enforce stricter compatibility standards. 

Conversely, training models exclusively on synthetic Expert Rules yields poor performance, emphasising that exposure to at least a portion of the historical co-purchase logs remains essential for effective recommendation. A compelling middle ground is offered by utilising synthetic datasets for pretraining followed by transaction-based finetuning, which maintains high relevance metrics while marginally increasing attribute consistency.

\subsection{Complementary Categories Mapping}
\label{sec:results:comcat}
Fundamentally, the ComCat mapping addresses two main challenges: the lack of explicit target categories in online environments, and the need to mitigate noise in historical co-purchase datasets (Section~\ref{sec:methods:dataset-construction}). To evaluate its effectiveness, an offline ablation study was conducted across four configurations of the mapping framework (Section~\ref{sec:methods:comcat}).

While the platform spans over 20,000 categories, results show that all configurations consistently cover approximately 42\% of source query categories. As shown in Table \ref{tab:offline-results-category-mapping}, while this source coverage remains stable, target category coverage increases significantly from 27.55\% in the baseline to 43.06\% in the final configuration. This confirms that adding more sources enriches the mapping with a deeper variety of high-quality target pairs without requiring a massive expansion of the source category set. The impact of this enrichment is most evident in the target distribution: while the median (p50) count of target categories remains stable at 3 or 4, the 95th percentile (p95) expands substantially from 3 to 10 categories as annotations, expert rules, and query categories are integrated.

To evaluate how these structural changes perform under realistic production conditions, an evaluation dataset was generated based on historical user traffic with known query items and engaged co-purchased or co-clicked categories. This setup enables the measurement of online-like CTR and CVR metrics in an offline experiment. Within this traffic-based dataset, the 42\% source coverage translates to 99.8\% of live traffic, confirming that the framework successfully provides at least one complementary target for virtually every relevant user interaction.
Consequently, these mappings prove highly effective in practice by covering the specific categories that drive the majority of user interactions. These quality gains are directly reflected in the final metrics: relative to the baseline (1), the full ensemble in (4) yields a +450\% increase in CTR and a +398\% boost in CVR. The substantial performance leap observed in (4) when including the same-category relation (Section \ref{sec:methods:comcat:merging}) suggests that, within the context of the analysed placement, users have a strong expectation for alternative product suggestions alongside complementary ones.

\begin{table*}
    \centering
    \Description[alttext]{alttext}
    \caption{Ablation Study of Category Mapping Sources based on offline traffic data from organic carousel on product page. Best results are bolded.}
\begin{tabular}{lcccccccccc}
    \toprule
    & \multicolumn{4}{c}{\textbf{Sources of category mapping}}                           & \multicolumn{2}{c}{\textbf{Taxonomy coverage $\uparrow$}} & \multicolumn{2}{l}{\textbf{Target counts $\uparrow$}} & \multicolumn{2}{c}{\textbf{Traffic metrics $\uparrow$}} \\
    \midrule
{No.}  & {Heuristics} & {Annotations} & {Rules} & {Same-Category} & {Query}         & {Targets}        & {p50}         & {p95}        & {CTR}          & {CVR}         \\
    \midrule
(1) & \ding{51} & & & & 41.77\% & 27.55\% & 3 & 3 & - & - \\
(2) & \ding{51} & \ding{51} & & & 41.89\% & 28.69\% & 3 & 5 & +25\% & +36\% \\
(3) & \ding{51} & \ding{51} & \ding{51} & & 42.32\% & 32.46\% & 3 & 9 & +125\% & +136\% \\
(4) & \ding{51} & \ding{51} & \ding{51} & \ding{51} & \textbf{42.32\%} & \textbf{43.06\%} & \textbf{4} & \textbf{10} & \textbf{+450\%} & \textbf{+398\%}   \\  
    \bottomrule
\end{tabular}
    \label{tab:offline-results-category-mapping}
\end{table*}

\subsection{Online A/B testing}
\label{sec:results:online-abtesting}

\begin{table}[t]
    \centering
    \small
    \caption{Summary of Online A/B Test Results. Metrics represent relative change against the production baseline. ($\ast$ denotes statistical significance at $p < 0.005$).}
    \begin{tabular}{clllcc}
        \toprule
        & & & & \multicolumn{2}{c}{\textbf{Platform}} \\
        \cmidrule(lr){5-6}
        \textbf{No.} & \textbf{Placement} & \textbf{Type} & \textbf{Metric} & \textbf{Web} $\uparrow$ & \textbf{App} $\uparrow$\\
        \midrule
        (1) & Product Page & Sponsored & v-CVR & +0.53\%$\ast$ & +0.13\% \\
        & & & GMV & +0.19\% & -0.01\% \\
        \midrule
        (2) & Product Page & Organic & v-CVR & +0.57\%       & -0.07\%\\
         & & & GMV & +8.05\%$\ast$ & +9.35\%$\ast$ \\
        \midrule
        (3) & Pre-Cart & Organic & v-CVR & -0.28\%$\ast$ & -0.04\% \\ 
        & & (filtered) & c-CVR & -0.51\% & -0.42\% \\ 
        & & & GMV & +0.17\% & -0.27\% \\
        (4) & Pre-Cart & Organic & v-CVR & -0.13\% & -0.17\% \\ 
        & & (finetuned) & c-CVR & +0.69\% & +0.35\% \\ 
        & & & GMV & -0.06\% & -0.28\% \\ \midrule
        (5) & In-Cart & Organic & c-CVR & +4.98\%$\ast$ & +1.21\% \\
        & & & GMV & +21.25\%$\ast$ & +15.73\%$\ast$ \\
        \bottomrule
    \end{tabular}
    \label{tab:online-results-abtests}
\end{table}
To validate the real-world efficacy of the AlleCompanion model, a series of online A/B tests was conducted across key platform placements. All organic carousels were limited to same-seller items, aligning with business logic intended to help users reach the free delivery threshold (MOV). Performance was measured primarily through Visit Conversion (v-CVR), representing the ratio of visits resulting in a purchase, Carousel Conversion (c-CVR), representing the ratio of carousel clicks resulting in a purchase, and the Gross Merchandise Value (GMV) directly linked to carousel interactions. A summary of these results is provided in Table \ref{tab:online-results-abtests}.

All A/B tests were performed on the web-facing version of Allegro (available on desktop and mobile web, referenced as Web) and the Allegro mobile app (referenced as App). Each experiment was conducted over a two-week period and routed 100\% of the platform traffic. AlleCompanion was evaluated as an additional retrieval source alongside existing production models.

\subsubsection{Product Page} 
Testing on the Product Page focused on two carousels --- Sponsored and Organic --- to evaluate how different recommendation strategies align with shifting user needs. In both placements, AlleCompanion was benchmarked against the existing production baseline, which combined item-to-item collaborative filtering on purchases, seller bestsellers, and seller new arrivals.

In the Sponsored placement (1), titled \textit{Suggestions for you}, AlleCompanion was evaluated utilising the ComCat mechanism, focusing exclusively on complements. This configuration yielded a +0.53\%$\ast$ and +0.13\% increase in v-CVR on Web and App, respectively. Although non-significant fluctuations were observed in GMV during the test, the carousel alone saw an approximate 50\% boost in ad revenue across both platforms, attributed to uplifts in CTR.

Conversely, the Organic carousel (2), titled \textit{Order in one shipment}, serves as a broader discovery tool. Initial evaluations suggested that purely complementary recommendations were overly restrictive for this placement, as users in this context often seek alternative products to enrich their selection alongside complements. By expanding the mapping with a same-category relation to support a ``\textit{complements + substitutes}'' combination (Section \ref{sec:methods:comcat:merging}), the system effectively broadened its discovery range. This was reflected by a boost in GMV of +9.35\%$\ast$ on App and +8.05\%$\ast$ on the Web, while v-CVR showed neutral fluctuations.

\subsubsection{Cart Placements} 
Cart placements appear during the final stages of the user journey, including the ``pre-cart'' pop-up window and the ``in-cart'' view during checkout. In these placements, AlleCompanion was evaluated against a production baseline consisting of item-to-item collaborative filtering on purchases, seller bestsellers, and recurring purchases. These placements often feature carousels with slogans such as \textit{Buy more from this seller to get free delivery}, explicitly prompting users towards the MOV threshold. Since these placements specifically target order value optimization near checkout, GMV serves as the primary metric, and our analysis focuses exclusively on this outcome.

Two different strategies were tested in the Pre-cart layer. The first (3) evaluated variants from the dataset construction experiments (Section \ref{sec:results:dataset-construction}) to assess whether functional compatibility is preferred over relevance measured in offline. Specifically, the (3) Filtered Transactions variant (selected for its high attribute consistency) and the (4) Pretrained + Finetuned Same-Seller variant (selected as a balanced middle ground) were evaluated. The second strategy utilised a base transactional model with a same-category mapping (Section \ref{sec:methods:comcat:merging}) to introduce alternative product suggestions. While most core business metrics across both strategies showed non-significant fluctuations in GMV and c-CVR across platforms, the (3) Filtered Transactions variant led to a statistically significant -0.28\%$\ast$ drop in v-CVR on Web. Overall, these findings suggest that neither the inclusion of alternatives nor the use of expert-filtered rules fully satisfied user needs for this specific intent.

Interestingly, the In-cart placement (5) revealed a different dynamic. The same-category mapping that yielded neutral results in the pre-cart layer performed exceptionally well during the final checkout stage. In carousels encouraging users to consolidate their shipments, the model achieved increases of +15.73\%$\ast$ and +21.25\%$\ast$ in GMV on App and Web, respectively.

Based on A/B tests (1), (2) and (5), the model has been deployed to all three placements. Each deployment was justified by a statistically significant improvement in at least one primary metric (CVR or GMV) on at least one platform, without negatively impacting the remaining metrics.

\section{Conclusions}
\label{sec:conclusions}
We present AlleCompanion, a universal production-scale retrieval framework for complementary item recommendations. To guide basket building effectively, our design separates item-level fit from category-level intent, combining explicit item-level constraints in the input feature space with category guidance imposed via category adapter. By decoupling the ComCat category mapping from the retrieval model, this architecture provides a flexible, production-friendly paradigm where category policies can be updated dynamically without expensive model retraining. 

Our online experiments reveal a crucial insight: while baseline behavioural filtering is necessary to clear out transaction outliers, the model benefits most from exposure to the full remaining distribution of co-purchase traffic, whereas enforcing strict, domain-specific restrictions degrades overall retrieval performance. Consequently, we decouple data precision from model training by shifting expert logic, LLM-based reasoning, and human feedback entirely into our ComCat mapping layer. Importantly, while designed for complementary recommendations, our production deployment revealed that the strongest business gains emerged when expanding the system to support a mixed related-product strategy (blending complements with same-category alternatives). This highlights that real-world checkout intents often favor broader discovery over strictly complementary options. Deployed at scale to serve over 20 million active users at product page and cart, AlleCompanion successfully bridges the gap between co-purchase and compatibility, delivering a +8-9\% GMV increase in organic discovery on the product page, +15-21\% GMV uplift in the cart, and a 50\% revenue boost in sponsored placements.

While AlleCompanion successfully addresses large-scale complementary retrieval, certain design choices introduce opportunities for further development. Firstly, because our framework relies on the ComCat mapping as the primary driver of complementary constraints, its precision is naturally bounded by the granularity of the underlying taxonomy, which can occasionally obscure item-level nuances or encounter coverage gaps in extreme cold-start categories. Although this trend is not currently visible in the traffic logs, as the framework successfully covers 99.8\% of active customer interactions, the architecture could be generalized to learn directly from raw product features. Secondly, the current deployment prioritizes broad item-to-item retrieval and interleaving to boost carousel diversity, leaving room to incorporate user personalization directly into candidate generation or via a dedicated downstream ranking layer. Finally, evaluating multi-item complementary intent in an offline manner remains difficult due to the inherent feedback loop from the production system. Consequently, online A/B testing remains the gold standard for reliably measuring true multi-item purchase dynamics.


\begin{acks}
This work is the result of a collaborative effort within the recommendation systems team at Allegro. We would like to extend our gratitude to fellow researchers Paweł Młyniec, Mateusz Marzec, and Bartłomiej Szołkowski for their invaluable support. We also thank the engineering team — Jakub Demianowski and Mateusz Lamecki — as well as former members Krzysztof Szczepański, Maciej Arciuch, Elwira Hołowko and Marcin Cylke for their foundational contributions to ML-based complementary recommendations at Allegro. Special thanks also go to the Paloma team behind the expert rule-based systems.
\end{acks}

\balance
\bibliographystyle{ACM-Reference-Format}
\bibliography{bibliography}

\end{document}